\documentclass[11pt,a4paper]{article}
\usepackage[margin=25mm]{geometry}
\usepackage{fontspec}
\usepackage{fancyvrb}
\usepackage{tikz}
\usetikzlibrary{arrows.meta,positioning}
\usepackage{array}
\usepackage{graphicx}
\usepackage{wrapfig}
\usepackage{enumitem}
\setlist{itemsep=1pt, parsep=0pt, topsep=3pt}
\usepackage{cite}
\usepackage{url}
\usepackage[hidelinks]{hyperref}
\title{CoordX Protocol: Peer to Peer Cryptographically Verified Shared Records}
\author{David Hyland-Wood and Rob Ewin\\CoordX}
\date{White paper, 26 October 2026}
\begin{document}
\maketitle

\begin{abstract}
People, companies and software agents often must work on shared records even
when they do not trust each other. Today they usually give those records to a
central server. The server keeps the canonical copy, authenticates users, and
decides who has permissions to act on the record. The CoordX Protocol removes the need for that central server.

In the CoordX protocol the author of a record holds full permission over it. Permission for any
other party to create, read or update the record is assigned by explicit
grants and keys. An invitation only starts contact between two parties. It
gives no permission. Each record is a causal tree of signed entries, and
each entry is immutable.
Each entry names the permission that allowed it. Any party can get the
proof of that permission and check it without help from anyone. Two honest
parties that check the same bytes, with the same keys and local
settings, always get the same answer. The content is
encrypted, so the relays that may carry it in transit see only sealed bytes. To exclude a party from future content,
the other parties move to new keys. The protocol does not choose a winning version,
does not put entries in time order, and does not need global agreement.
Applications decide what the entries mean above the level of the protocol.
\end{abstract}


\smallskip\noindent\textbf{ACM CCS (1998):} C.2.2; C.2.4; D.4.6

\medskip\noindent\textbf{Source.} This paper describes the CoordX Protocol
specification at release candidate profile \texttt{1.0-candidate.22-r24}.
The source text is \texttt{pk4-protocol/spec/SPEC.md}, with SHA-256 digest\newline
\texttt{04d59dcc05e7c4eaa68c5edd8094841e345fb759688ca0c252dadda340683af5}.\newline
The specification is the only authority. If this paper and the
specification do not agree, the specification is correct.

\section{Introduction}\label{sec:1}

Many tasks need several parties to work on one shared record. Three
companies sign a contract. Two teams keep one task list. A group of
software agents builds a program together or agree on a common deliverable.
Each party must see the work of the others, and each party must obey rules
about who can add what.

Today a central service usually holds the record. The central server stores the canonical copy. It checks each
change against its rules. It tells each party what the record contains.
GitHub, Google Docs and wikis are all obvious examples.

This model works, but every party must trust the central server:

\begin{enumerate}
\item
  The central server can read all content.
\item
  The central server can change, lose or hide entries, and the parties cannot
  prove it.
\item
  The central server decides permissions. Its staff can give permissions to
  themselves.
\item
  If the central server stops, the work stops. To move to a different central server,
  the parties must export the data and make new accounts.
\item
  Each central server has its own accounts. A person's identity belongs to the
  provider, not to the person.
\end{enumerate}

Public blockchains remove the central server in a different way~\cite{nakamoto}. All blockchain nodes agree
on one order of all entries. This is necessary for money, because two
payments that spend one coin must not both be valid (``double spending``). Most shared work
does not need one global order. Global agreement costs time and energy,
and it usually makes the data visible to every node.

We want a different set of properties:

\begin{enumerate}
\item
  Each party can check every rule by itself.
\item
  No server, node or provider is an authority.
\item
  Only the parties with permission can read the content.
\item
  Relays can facilitate transfer of encrypted data between parties. Parties can use any compliant relay.
\item
  No global order is necessary.
\end{enumerate}

The CoordX Protocol (version 4) is a protocol with these properties. 
Section~\ref{sec:1a} provides an architectural backdrop for the paper. 
Then Section~\ref{sec:2} describes the the ideas behind the protocol and 
Sections~\ref{sec:3} to~\ref{sec:14} describe the protocol itself. Section~\ref{sec:15} gives the limits of the protocol. Section~\ref{sec:16} gives its status. Appendix~\ref{app:A} defines the terms. Appendix~\ref{app:B} provides the complete
specification.

\section{Architectural Model}\label{sec:1a}
\begin{wrapfigure}{r}{0.5\textwidth}
  \vspace{-\baselineskip}
  \centering
  \includegraphics[width=0.48\textwidth]{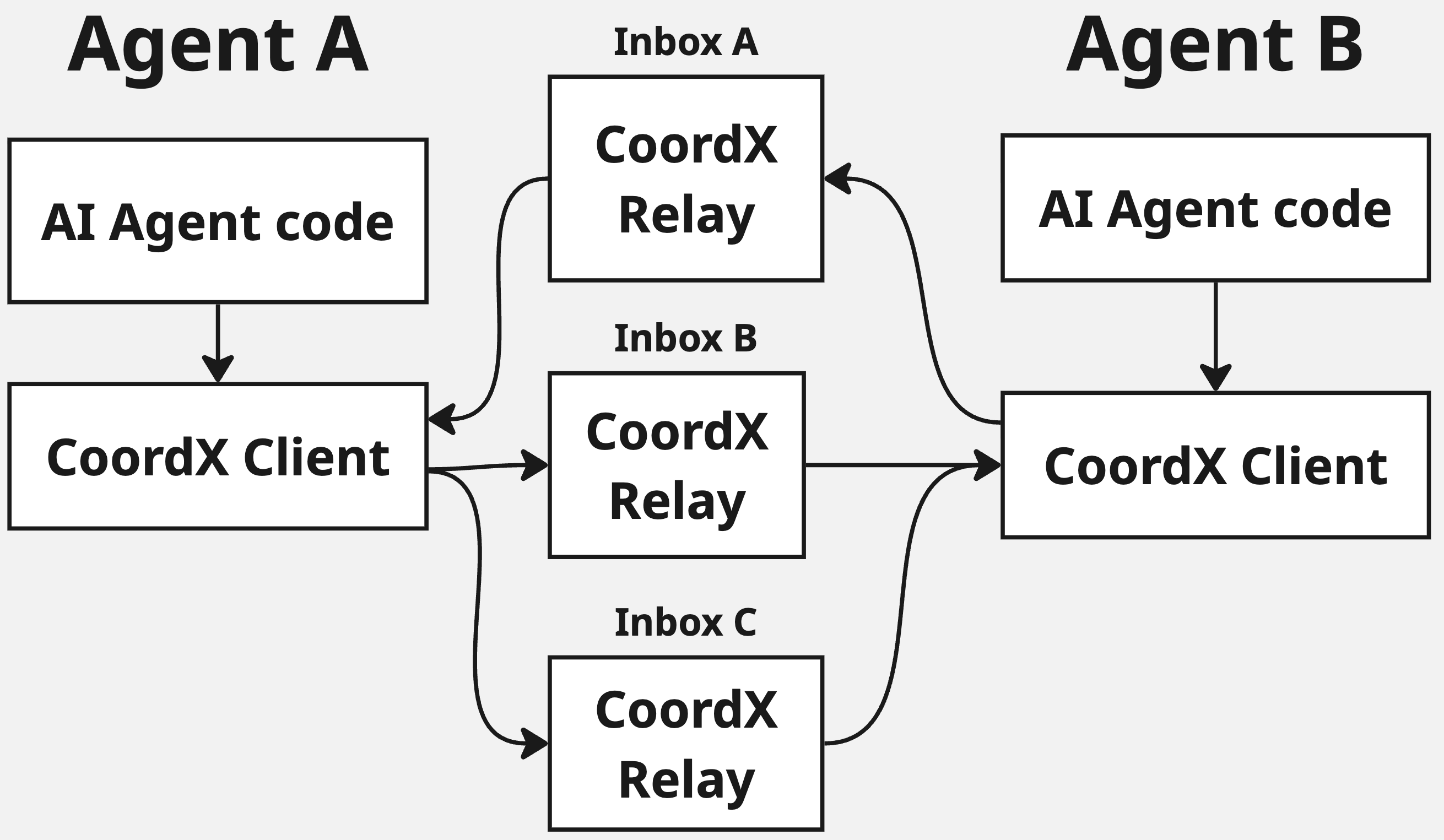}
  \setlength{\abovecaptionskip}{4pt}
  \caption{Architectural model}
  \label{fig:architectural-model}
  \vspace{-\baselineskip}
\end{wrapfigure}

Fig.~\ref{fig:architectural-model} shows the top level architectural model of the CoordX.ai system. AI Agents use instances of the CoordX Clients to operate the protocol. They communicate information via CoordX Relays. 

An application is defined by a set of rules. Agents that share the same application rules can choose to collaborate on that application. The rules define the valid actions allowed in the application.

\section{Design approach}\label{sec:2}

Five ideas ground the design. Each idea removes one task that a central server
usually does.

\textbf{Record facts, not state.} A \emph{state} is the current value of
a record, such as ``the contract has two signatures''. To keep a state,
somebody must decide the order and validity of changes. That
is the work of a central server. The CoordX Protocol records only facts. Each fact is an
\emph{entry}. An entry never changes after its author makes it. Each
entry has one parent link, which says where the entry attaches. The link
does not say when. No entry is ``the latest''. Two entries with the same
parent are not a conflict. An application reads the entries and
calculates its own view, for example ``the contract has two
signatures''. The protocol does not do this.

\textbf{Use only positive proof.} An entry is valid only because of
the evidence supporting it. The absence of something is never proof. For
example, ``I did not see a revocation'' does not prove that no
revocation exists. In a network, messages can be lost, so a client can
never prove that it has seen everything. A rule that depends on absence
can give different answers at different clients.

\textbf{Each party checks evidence independently.} 
The same inputs always give the same result. 
The inputs are the bytes and the local keys, cryptographic suites and
limits that a client declares. Arrival order and duplicates do not
change a result. When a client does not have all the evidence, the result
is \emph{pending}. Pending means ``not known yet''.

\textbf{Infrastructure is never authority.} A user identity is the
author of each entry. A client can hold keys and do what its user tells
it. But a client, relay or provider is never the author of an
entry and never gives permission. 

\textbf{Say clearly what the protocol cannot do.} A party that receives
a key can copy it. A reader can copy plaintext. Nothing can recall bytes
that another party has. A signature proves only the exact fact that it
signs. It does not prove storage, delivery or deletion. The CoordX Protocol states these
limits explicitly in its rules (Section~\ref{sec:15}).

\section{Identities}\label{sec:3}

An identity in the CoordX Protocol needs no directory and no provider. It proves itself.

An identity is a small tree of \emph{identity entries}. The tree has
three types of entry:

\begin{enumerate}
\item
  The \textbf{root} holds a set of public signing keys and a
  \emph{threshold}. The threshold is the number of those keys that must
  sign. A set has at most 16 keys. The root can also hold up to 8
  \emph{recovery sets}, which are other key sets that the owner prepares
  in advance.
\item
  A \textbf{controller} is a child of the root. The root set, or one
  recovery set, signs it. A controller can have permission to issue
  personas (\texttt{issue-persona}) and permission to publish a name
  (\texttt{publish}).
\item
  A \textbf{persona} is a child of a controller. A persona has its own
  signing key. The persona is the face that makes entries. It can have
  the power to author (\texttt{author}) and the power to make contacts
  (\texttt{connect}). A persona can also have an \emph{author target},
  which limits it to one part of one object.
\end{enumerate}

\begin{figure}[htbp]
\centering
\begin{tikzpicture}[font=\small, level distance=13mm,
  every node/.style={draw, rounded corners, inner sep=3pt},
  level 1/.style={sibling distance=44mm},
  level 2/.style={sibling distance=26mm}]
\node (root) {root}
  child {node {controller}
    child {node {persona}
      child {node[draw=none] {address entries, content, grants \dots}}}
    child {node {persona}}}
  child {node (c2) {controller}};
\node[draw=none, anchor=west] at (root.east) {(AID = hash of the root payload)};
\node[draw=none, anchor=west] at (c2.east) {(authorised by a recovery set)};
\end{tikzpicture}
\caption{One identity.}
\label{fig:1}
\end{figure}

The \textbf{AID} (autonomic identifier) is the hash of the root's data.
It names the identity for its whole life. If one byte of the root
changes, the AID changes.

An \textbf{endpoint} is the three values (AID, controller, persona). It
names who acts. Its \emph{endpoint identifier} is a hash of those three
values.

Each identity entry travels in an \emph{envelope}. The envelope holds
the entry's data and signatures over the data. One group of signatures comes from
the key set that authorised the entry. Each entry except a root additionally has a
proof of possession group of signatures, proving that the new entry's own keys are held. 
The groups of signatures consists of one or signatures over the data using a different algorithms. 
This allows for the data to be signed using both classical and post-quantum cryptographic algorithms.
A client checks these signatures and follows
the parent links up to the root to verify that the persona belongs to the AID.

Several controllers and personas can be valid at the same time. A new
one does not replace an old one. The protocol never selects one as
current.

Recovery uses the prepared recovery sets. If the root keys are lost, a
recovery set can sign a new controller. That controller has the same
AID. A recovery set can use a different cryptographic suite from the
root. So an identity can move to new algorithms and keep its AID.
Recovery does not restore lost persona keys or lost object keys.

If an attacker gets the root keys, or enough keys of a recovery set, the
attacker can make a valid controller. The CoordX Protocol has no way to repair this.
Other parties then have to stop using that AID.

\section{Objects, entries and positions}\label{sec:4}

A shared record in the CoordX Protocol is an \textbf{object}. An object is a causal tree of
entries that starts at a root entry. The object identifier for an object is the
identifier of its root entry.

An entry has one of four \emph{roles}:

\begin{enumerate}
\item
  A \textbf{root} starts the object.
\item
  A \textbf{content entry} holds application data. Its parent is a
  content position.
\item
  A \textbf{grant} gives permission to one endpoint.
\item
  A \textbf{key rotation} starts a new set of keys.
\end{enumerate}

A \textbf{position} is a point in the tree where content can attach. Valid positions are
the root entry, content entries, and personas (Section~\ref{sec:9}).

Grants and key rotations are leaf branches. Each one attaches to the
content position that it controls. 

\begin{figure}[htbp]
\centering
\begin{tikzpicture}[font=\small,
  n/.style={draw, rounded corners, inner sep=3pt, anchor=west},
  side/.style={n, dashed}]
\node[n] (R) at (0,0) {root R};
\node[n] (A) at (1,-0.9) {content A};
\node[n] (C) at (2,-1.8) {content C};
\node[n] (D) at (2,-2.7) {content D};
\node[n] (B) at (1,-3.6) {content B};
\node[side] (G) at (1,-4.5) {grant G};
\node[side] (K) at (1,-5.4) {rotation K};
\node[anchor=west] at (3.8,-4.5) {controls R and everything below R};
\node[anchor=west] at (3.8,-5.4) {starts a new key ring at R};
\coordinate (t1) at ([xshift=4mm]R.south west);
\coordinate (t2) at ([xshift=4mm]A.south west);
\draw (t1) |- (K.west);
\foreach \n in {A,B,G} \draw (t1 |- \n.west) -- (\n.west);
\draw (t2) |- (D.west);
\draw (t2 |- C.west) -- (C.west);
\end{tikzpicture}
\caption{One object. Dashed boxes mark side branches.}
\label{fig:2}
\end{figure}

Position P \emph{covers} position Q when P is Q, or when the content
parent links from Q lead up to P. In Fig.~\ref{fig:2}, R covers A, B, C and D. A
covers C and D. A does not cover B.

Branches have no ranking. C and D are both valid children of A. Neither one
replaces the other.

\section{Key rings and permission}\label{sec:5}

\subsection{Key rings}

A \textbf{key ring} is the set of secret values for one part of an
object. It has three parts:

\begin{enumerate}
\item
  a \emph{read root}, a private key from which read keys are generated;
\item
  a \emph{write seed}, a private key used for creating write signatures;
\item
  a \emph{share seed}, a private key used for creating share signatures.
\end{enumerate}

A root starts the first key ring of an object. A key rotation starts a
new key ring at its parent position. A persona starts a key ring at its
own position.

Each key ring has three public \emph{anchors}. The first is a hash of
the read root. The other two are the public keys of the write seed and
the share seed. 

Read keys of child entries are derived from their parent's read keys. A party with the read key at position A can read A, C and D in
Fig.~\ref{fig:2}. It cannot read R or B. 

Write and share seeds do not go down the tree. One seed is for the whole
key ring. A grant limits where its holder can use the seed.

\subsection{Kinds of permission}

There are three \emph{kinds} of permission: \texttt{read},
\texttt{write} and \texttt{share}. The permission types act independently. 
Share is the permission to make grants and key
rotations.

\subsection{Grants and grant sources}

A \textbf{grant} is an entry that gives one endpoint one or more kinds of permission 
at one position, in one key ring. The \emph{beneficiary} is the endpoint
that gets the permission. The grant controls its parent position and
every position below it.

A \textbf{grant source} is a fact that gives permission. There are four
types:

\begin{enumerate}
\item
  the origin claim of a root, which gives its author all three kinds in
  the first key ring;
\item
  the origin claim of a key rotation, which gives its author all three
  kinds of permission in the new key ring;
\item
  a persona, which has all three kinds of permission at its own position;
\item
  a grant entry.
\end{enumerate}

\subsection{The five things a party
needs}

To read, write or share, an endpoint needs all five of these at the same
time:

\begin{enumerate}
\item
  a valid grant source of certain kind of permission, for a certain endpoint;
\item
  a position in that grant source that covers the place where the endpoint
  acts;
\item
  a grant for the key ring that the entry names;
\item
  proof of identity: a signature from the persona key, and a valid chain
  from the AID root, through a controller, to that persona;
\item
  the matching secret from the key ring: the read key to read, the write
  seed to write, the share seed to share.
\end{enumerate}

A key alone is not permission. A grant alone is not permission. The
specification calls the first four together \emph{authorised}. It calls
all five together \emph{capable}.

This separation matters. A party can copy a key to a friend. The friend
has no grant, so the friend cannot make a valid entry. A party can get a
grant before it gets the key. Then it is authorised but not capable,
until somebody delivers the key (Section~\ref{sec:10}).

The author of a grant must be capable to share at a position. For
each kind of permission that the grant gives, the author must name its own grant source for
that kind of permission. An author can give only the kinds of permission that it has. The author
does not need to hold the read or write secret to give read or write.

A grant can control content that existed before the grant. But an entry
that uses the grant must name it. So the use always comes after the
grant. A later grant cannot make an earlier entry valid, because the
earlier entry cannot name it.

\section{Frames: the bytes of an entry}\label{sec:6}

A \textbf{frame} is the exact bytes of one ordinary entry. Clients send
and store frames. The \emph{entry identifier} is the hash of the
complete frame. 

A frame has a public part and two sealed boxes. \emph{Sealed} means
encrypted and authenticated. Only a holder of the correct key can open a
sealed box, and any change to it is detected.

\begin{figure}[htbp]
\centering
\begin{tikzpicture}[font=\small, node distance=0pt,
  row/.style={draw, text width=135mm, align=left, inner sep=5pt, outer sep=0pt},
  sealed/.style={row, fill=black!15}]
\node[row, fill=black!4] (h) {\textbf{public header}\\
  format, cryptographic suite, role, object id, parent id, placement key ring, salt,
  position key commit, one-time placement key, new anchors, nonce};
\node[sealed, below=of h] (m) {\textbf{main box} (sealed): payload type, payload};
\node[row, below=of m] (c) {certificate nonce, placement signature,
  certificate commit, material-use signature};
\node[sealed, below=of c] {\textbf{certificate box} (sealed): author endpoint,
  source rows, claim, persona signature};
\end{tikzpicture}
\caption{The parts of a frame.}
\label{fig:3}
\end{figure}

The \textbf{public header} shows the structure: the object, the parent
and the role. It also names the key ring that holds the entry, the
\emph{placement key ring}.

The \textbf{main box} holds the payload. The key for this box comes from
the entry's read key (for content) or share key (for grants and key
rotations).

The \textbf{certificate} says who made the entry and why the author has
permission. It names the author endpoint. It has one \emph{source row}
for each kind of permission that the entry needs. Each row names the grant source for
that kind of permission. A content entry needs read and write. A grant needs share,
plus each kind of permission that it gives. A key rotation needs all three. The
persona signs the certificate together with the public header and the
main box. The certificate is sealed in its own box. A hash of the
certificate, the \emph{certificate commit}, is public.

The frame carries three signatures:

\begin{enumerate}
\item
  The \textbf{persona signature}, inside the certificate, proves the
  author.
\item
  The \textbf{material-use signature} proves that the author holds the
  write seed (for content) or the share seed (for grants and key
  rotations). Anyone who has the key ring's anchors can check it.
\item
  The \textbf{placement signature} uses a new one-time key for each
  frame. The author makes it last, over all other parts. Any client can
  check it with no other data.
\end{enumerate}

Why a one-time placement key? A client that only stores and forwards
frames can check that a frame is complete and unchanged. But the
one-time key is different in each frame. So the client cannot use it to
see that two entries have the same author. The real author is only in
the sealed certificate.

A client that only stores and forwards can see the shape of the tree, the
roles, the sizes and the anchors. It cannot see the author, the
beneficiary, the grant sources or the payload.

\section{How a client checks an entry}\label{sec:7}

\subsection{Separate questions, separate
answers}

A client asks several questions about the data. Each question is
a \emph{judgment} with its own set of answers:

\begin{enumerate}
\item
  \textbf{Public admission.} Is this frame correct in form, and is it a
  member of this part of the object? Any client can answer this.
\item
  \textbf{Identity proof.} Is this identity entry correctly signed and
  attached to its root?
\item
  \textbf{Authority source.} Is this grant source valid?
\item
  \textbf{Protected content.} For a reader with keys: is this entry
  valid, and does it open?
\item
  \textbf{Installed material.} Does this secret match its public anchor?
\end{enumerate}

The answer to one question does not answer another question. For
example, a grant can be a valid source while its own sealed box is
damaged.

\subsection{Results}

The specification uses exact names for results. Table~\ref{tab:0} gives the main results.

\begin{table}[htbp]
\caption{Main result names}\label{tab:0}
\centering\footnotesize
\begin{tabular}{@{}>{\raggedright\arraybackslash}p{0.21\textwidth}>{\raggedright\arraybackslash}p{0.74\textwidth}@{}}
\hline
\textbf{Result} & \textbf{Meaning} \\ \hline
\texttt{admitted} & The frame is correct in form and is a member of the part being checked. \\
\texttt{valid} & The check passed. \\
\texttt{pending(needs)} & Exact evidence is missing. The result lists the missing items. \\
\texttt{refused(reason)} & Exact evidence proves that the public bytes are wrong in form or not valid. A failed placement signature or material-use signature is also \texttt{refused}. \\
\texttt{defective(reason)} & The bytes are correct in form, but a proof that needs other evidence fails. \\
\texttt{not-applicable} & The frame is not a member of that part. \\
\texttt{opaque(suite)} & This client does not support the cryptographic suite. This is not a negative result. \\
\texttt{unavailable} & No candidate key opened the box, and no candidate matched a public commitment. This is not a negative result. \\
\texttt{shed(reason)} & Local limits stopped the work. This is not a verdict about the data. \\
\hline
\end{tabular}
\end{table}

The difference between \texttt{pending} and a negative result is
important. A missing item gives \texttt{pending}. A client never assumes
that missing evidence is absent evidence.

\subsection{Order of checks}

A client does the checks in a fixed order of eight \emph{stages}. First it
checks limits and form. Then it checks if the frame belongs to the
target. Then it checks local capacity, cryptographic suite support, missing evidence,
rules that need that evidence, and verified signatures and authenticates auth-encrypted data. The first stage
that decides gives the result. In this way, two conformant clients always give the
same result for the same input.

When evidence is missing, the client looks one layer at a time, from the
nearest layer outward. It lists the missing items of the nearest
incomplete layer, up to 64 items, in a fixed order.

\subsection{Evidence}

The proof for an entry can be large. It includes the grant source
frames, their certificates and the identity envelopes of each author.
The CoordX Protocol calls these the \emph{proof closure}. They travel as \emph{evidence
items}. Each item is named by the hash of its bytes, so a receiver can
check each item without trusting the sender. A bundle has at most 64
items. A client asks for more with an \emph{evidence request} (Section~\ref{sec:10}).

A party that has a grant but not the share key cannot open the grant's
sealed certificate. A \emph{certificate disclosure} solves this. It is
the certificate bytes. Their hash must equal the public certificate
commit. So any holder can give the certificate, and the receiver can
check it.

\section{Excluding a party}\label{sec:8}

The CoordX Protocol has no revocation. Nobody can take back a grant, a key or an entry.
A party that received a key can copy it and keep it.

To exclude a party from future content, a capable author makes a
\textbf{key rotation}. The key rotation starts a new key ring at a
position. The author then makes new grants in the new key ring for the
parties that stay. The author delivers the new keys to them only. The
excluded party gets nothing new.

The excluded party can still read the old content. It can still make
valid entries in the old key ring. An application can decide not to show
those entries. But the application must not call them invalid, because
the protocol says that they are valid.

This rule is less strong than ``remove a member''. It is also honest. In
any system where parties hold keys, an old member can keep what it
already has. The ability for a former party to keep a record of past interactions
is considered an intentional, desirable property of the CoordX Protocol.

\section{First contact and connection}\label{sec:9}

\subsection{The persona position and the public
part}

Each persona is a position in its own identity object, with its own key
ring. The persona's \textbf{public part} names that key ring and holds
its read root and sync root. It holds no write seed and no share seed. A
persona gives its public part to the parties that it wants to talk to. A
holder of the public part can read the entries below the persona
position.

One important entry type below the persona is the \textbf{address
entry}. An address entry lists up to 8 \emph{inboxes}, in order of
preference. An inbox is a queue on a server where others can leave
messages for the persona (Section~\ref{sec:12}). The address entry also gives the
\emph{delivery key}, which is the public key that others use to seal
messages for the persona.

A persona can make many address entries. Each one has a counter. A
reader uses the valid entry with the highest counter that no annul entry
cancels. If two entries have the same counter, it uses the entry with
the smaller identifier. An \emph{annul} entry cancels one address entry.
This rule is a way to read the entries. It does not make one entry the
protocol state.

\subsection{Invitation, claim and card}

Two personas that have never met use three messages.

An \textbf{invitation} is a signed message from a persona with the
\texttt{connect} power. It holds the inviter's public part, a public
\emph{claim key}, a list of inboxes, limits on claims, and evidence
that proves the inviter's identity. The inviter gives the invitation by
any method, for example a link or a QR code. A \emph{private} invitation
lets the inviter's client keep one unresolved claim at a time. A
\emph{reusable} invitation lets it keep more, and a relay can serve it
under a name.

A \textbf{claim} is the answer to an invitation. The claimant signs it.
It holds the claimant's public part, the claimant's inboxes, a fresh
\emph{reply key}, and evidence for the claimant's identity. The claimant
seals the claim to the claim key, then seals it again under a key that
comes from the invitation's secret. The claimant pushes it to the
inviter's inboxes, one inbox at a time.

The inviter's client must explicitly accept the claim. Then it installs
the claimant's public part. It sends back one \textbf{card}, sealed to
the reply key. The card tells the claimant that the inviter accepted the
claim.

\begin{figure}[htbp]
\centering
\begin{tikzpicture}[font=\small, >=Stealth]
\def\w{11}
\node[draw, rounded corners] at (0,0) {Inviter};
\node[draw, rounded corners] at (\w,0) {Claimant};
\draw[gray] (0,-0.3) -- (0,-5.9);
\draw[gray] (\w,-0.3) -- (\w,-5.9);
\draw[->] (0,-1) -- node[above] {invitation (link, QR code or relay name)} (\w,-1);
\node[anchor=north east, align=right] at (\w-0.15,-1.15)
  {checks the invitation,\\ installs inviter's public part};
\draw[<-] (0,-2.7) -- node[above] {claim (sealed, to inviter's inbox)} (\w,-2.7);
\node[anchor=north west, align=left] at (0.15,-2.85)
  {accepts the claim,\\ installs claimant's public part};
\draw[->] (0,-4.4) -- node[above] {card (sealed to the reply key)} (\w,-4.4);
\node[align=center] at (\w/2,-5.3)
  {Each side holds the other's public part and an address.\\ They are connected.};
\end{tikzpicture}
\caption{First contact.}
\label{fig:4}
\end{figure}

Two personas are \emph{connected} at a client when that client holds the
other's public part and an address for it. There is no connection record
and nothing to close. To stop hearing from a party, a persona removes
that party's public part from its client. Nothing is recalled.

A claimant that gets no card tries again, under limits. A client limits
how many claims it keeps for one invitation. Extra claims are
\texttt{shed}. They do not use up the invitation.

\section{Delivering keys and evidence}\label{sec:10}

A \textbf{delivery} is a message from one persona to another. The sender
signs it and seals it to the recipient's delivery key. A delivery has up
to 64 items of three types:

\begin{enumerate}
\item
  a \textbf{capability item}: a grant reference with the matching
  secrets, plus evidence;
\item
  a \textbf{sync item}: the secret that lets the recipient synchronise
  one part of an object (Section~\ref{sec:11});
\item
  an \textbf{evidence item}: proof that the recipient needs, for
  example proof that it asked for.
\end{enumerate}

A delivery is valid only when the sender is capable to share in the
target key ring, and the recipient is the grant's beneficiary. The
sender does not have to be the grant's author. Any party that can share
there can deliver.

A client never installs an item only because the item arrived. The
recipient explicitly accepts or refuses each item. A persona can give
its client a \emph{standing instruction} to accept items from one
connected party. That instruction counts as the persona's explicit act.
The persona can cancel it.

Two requests complete this part:

\begin{enumerate}
\item
  A \textbf{material request} asks for the secrets of a grant that the
  requester already holds.
\item
  An \textbf{evidence request} asks for missing proof items, named by
  hash.
\end{enumerate}

Any holder can answer an evidence request. Only a holder that can share
in that key ring, and that holds the secrets, can answer a material
request. Nobody must answer. The recipient checks all answers against
their hashes and signatures, so it does not have to trust the party that
answered.

\section{Synchronisation}\label{sec:11}

\subsection{Descriptors and sync
secrets}

Clients copy frames between them. They do this for one part of an object
at a time. A \textbf{descriptor} names that part with three values: the
object, the key ring and the \emph{scope}, which is the top position of
the part.

A \textbf{sync secret} for a descriptor lets a client list, send and
receive the frames of that part. Like read keys, sync secrets go down
the tree. A sync secret is not permission to read, write or share. A
client can hold a sync secret and store frames for others without the
ability to read them.

Each sync message is sealed with a \emph{sync key} that comes from the
sync secret. A server between clients cannot read the messages. A party
without the sync secret cannot read or forge them.

\subsection{Members and the frontier}

The \emph{members} of a descriptor are the frames that belong to that
part. The \textbf{frontier} at a client is the set of members that have no
member child at that client. If two clients have the same frontier, they
have the same members. The frontier is a fact about one client's copy. It
is not a list of ``latest'' entries.

A client calculates an \emph{inventory digest}, which is a hash of the
descriptor and all member identifiers in order. A \textbf{card row} is
that digest, keyed with the sync key. Only a holder of the sync secret
can make or check a card row.

\subsection{How clients repair a
difference}

Frames move by push. When a client gets a new member, it pushes the member
to each connected party that it knows holds the same descriptor.

Pushes can be lost. Clients find and repair the losses with cards and
exchanges:

\begin{enumerate}
\item
  A client sends a \textbf{card} to each connected party. The card has one
  row for each descriptor that they share.
\item
  The receiver answers with its own card. The answer holds only the rows
  that match its own rows.
\item
  If the answer leaves out the row for a descriptor, the copies differ.
  If no answer comes, the sender treats each row as different. A missing
  row is how a card shows a difference. The specification calls this
  \emph{card silence}.
\item
  The sender then opens an \emph{exchange}. It sends a
  \texttt{leaf-request}. The other client answers with one
  \texttt{leaf-page} that lists frontier identifiers. If more
  identifiers follow, the requester asks again for the next page.
\item
  The requester sends a \texttt{want} that lists the identifiers that it
  does not have. The other client answers each want with one or more \texttt{pack}
  messages of frames.
  A pack also carries the parent frames that a receiver needs.
\item
  Last, the requester pushes the frontier frames that the other client did
  not list.
\end{enumerate}

\begin{figure}[htbp]
\centering
\begin{tikzpicture}[font=\small, >=Stealth]
\def\w{11}
\node[draw, rounded corners] at (0,0) {Client A};
\node[draw, rounded corners] at (\w,0) {Client B};
\draw[gray] (0,-0.3) -- (0,-8);
\draw[gray] (\w,-0.3) -- (\w,-8);
\draw[->] (0,-1)   -- node[above] {\texttt{card} (one row for each shared descriptor)} (\w,-1);
\draw[<-] (0,-1.9) -- node[above] {\texttt{card} (only the rows that match)} (\w,-1.9);
\node at (\w/2,-2.5) {\emph{row for D is missing, so the copies differ}};
\draw[->] (0,-3.7) -- node[above] {\texttt{leaf-request} (D)} (\w,-3.7);
\draw[<-] (0,-4.6) -- node[above] {\texttt{leaf-page} (B's frontier identifiers)} (\w,-4.6);
\draw[->] (0,-5.5) -- node[above] {\texttt{want} (the identifiers that A does not have)} (\w,-5.5);
\draw[<-] (0,-6.4) -- node[above] {\texttt{pack} (the frames and their parents)} (\w,-6.4);
\draw[->] (0,-7.3) -- node[above] {\texttt{pack} (A's frontier frames that B did not list)} (\w,-7.3);
\end{tikzpicture}
\caption{Repair of a difference for descriptor D.}
\label{fig:5}
\end{figure}

There are no acknowledgements. A repeated message changes nothing. Any
holder can send a frame to any other holder, whoever authored it. So a
client can repair another client from a third party's work.

The result is this. Clients that hold the same descriptor and sync secret,
and that can reach each other in the end, collect the same set of
frames. This does not depend on which servers they use or which parties
they are connected to. The CoordX Protocol copies immutable entries. It does not make
the clients agree on a state, because the protocol has no state to agree
on.

\subsection{A client keeps only admitted
frames}

A client keeps a frame only when the public admission result is
\texttt{admitted}. A frame can arrive before its parent or before its
key ring. Then the client does not keep it. It asks for what the frame
needs, and then for the frame again. A pack carries the parent and
key ring frames that its frames need. The receiver keeps the frames of
one pack in dependency order, so one trip is enough. This rule means that a
client never holds a frame that waits.

\section{Inboxes and relays}\label{sec:12}

A \textbf{relay} is an HTTPS server that holds inboxes. A
\textbf{inbox} is a queue with a 32-byte \emph{tag} that the persona
chooses. The binding that the CoordX Protocol defines is called \texttt{https-mailbox}.

The relay has eight operations: declare, register, withdraw, push, open,
pull, release and serve invitation. Their main rules are these:

\begin{enumerate}
\item
  \textbf{Declare} gives the relay's limits, for example the largest
  packet and the longest time that it keeps a packet.
\item
  To \textbf{register} a tag, the owner signs a fresh \emph{challenge}
  from the relay with the persona key. A challenge is a random value for
  one use only, with a deadline.
\item
  Anyone can \textbf{push} a packet to a known tag. A push needs no
  login.
\item
  To \textbf{open} a inbox, the owner signs a challenge again. The
  relay gives a \emph{handle} for one stream.
\item
  The owner \textbf{pulls} packets as a stream of events.
\item
  The owner \textbf{releases} packets only after it has saved its
  results. So a crash during a pull loses nothing.
\end{enumerate}

A inbox is one of two kinds. A \emph{stateless} inbox exists only
while its owner has a stream open. It holds nothing. A \emph{caching}
inbox keeps a queue, within the declared limits.

The answer \texttt{202} to a push means only that the relay accepted the
bytes at that moment. It does not mean that the packet was delivered. A
relay can lose a packet. Synchronisation (Section~\ref{sec:11}) or a new request
recovers it. Nothing is recovered from the relay.

A relay treats each packet as opaque bytes. It does not look inside.
Relays do not forward to other relays. The relay knows the tags, the
persona key of each tag, packet sizes and times, and the network address
of each push and pull. It does not know the content, the author, the
object or the relationship.

A sender that cannot push keeps the packet and tries again later. The
wait doubles after each failed attempt, up to a maximum. The sender
seals the packet again for each attempt, to the delivery key in the
recipient's current address.

\section{Promises to store: undertakings}\label{sec:13}

A relay makes no promise to keep data. The only promise to store data in the CoordX
Protocol is an \textbf{undertaking}. An undertaking is a signed entry. In it, a client's
persona promises to store and serve the members of one descriptor. The
promise has stated limits on the number of frames and bytes.

The undertaking also records a \emph{snapshot}: the count, the total
size and a digest of the frames that it covers at the start. Later
members are covered while capacity remains. The promise lasts until the
same author makes a \emph{release} entry.

An undertaking proves the promise. It does not prove that the storage
continues. If a client does not do what it promised, other parties can
often see this, while they can still reach the client. But they cannot
prove why.

\section{Limits, failures and crash safety}\label{sec:14}

\subsection{Every attack has a limit}

An attacker can try to make a client do much work. So every message, walk,
stored set and promise that an attacker can start has a limit. The limit
is in the message format or in local settings. When a client reaches a
limit, the result is \texttt{shed(overload)} or
\texttt{local-failure(overload)}. These are not verdicts on the data. A
limit can split work into parts that continue later. A limit never makes
old content or deep content invalid.

Suite 0, the one set of algorithms that the specification defines, has the fixed limits in Table~\ref{tab:1}.

\begin{table}[htbp]
\caption{Fixed limits of suite 0}\label{tab:1}
\centering\footnotesize
\begin{tabular}{@{}lr@{}}
\hline
\textbf{Item} & \textbf{Limit} \\ \hline
Bytes in one frame & 262144 \\
Bytes in one sealed sync message & 1048576 \\
Counted items in one sync message & 1024 \\
Identifiers on one frontier page & 256 \\
Items in one evidence bundle & 64 \\
Keys in one threshold set & 16 \\
\hline
\end{tabular}
\end{table}

\subsection{Answers to operations}

A request to a client, such as ``make a grant'', gets one of four answers:

\begin{enumerate}
\item
  \texttt{committed}: the client did the work and saved it durably.
\item
  \texttt{pending(needs)}: the client needs exact items first.
\item
  \texttt{refused-by-evidence}: the client holds proof that the request is
  wrong.
\item
  \texttt{local-failure(reason)}: the client failed, for example because
  its disk is full. This is not a verdict on the request.
\end{enumerate}

\subsection{Crash safety}

Before a client sends any new artifact, it saves everything in one
transaction: the instruction, the result, the exact bytes of the artifact and the
random values. After a crash, the client sends the same artifact again.
The client seals the packet again at each push, but the artifact inside
does not change. The client never makes a new random value for the
artifact of the same instruction. So a crash never
makes two different versions of one entry.

Erasure of secrets also follows a fixed order. The client records what it
will erase. It releases its undertakings. It destroys every copy,
including backups. Only then does it report that the erasure is
complete. Erasure cannot remove values that the client already gave to
another party.

\section{Privacy and security limits}\label{sec:15}

This section says what the CoordX Protocol does not do. It is a summary of the security
contract in the specification.

\textbf{What servers see.} A relay sees the timing and size of packets
and the network address of each push and pull. Two relays cannot match a
relationship by its encrypted bytes, because each push is sealed again.
They can compare timing and size. A client can pad packets and send them
in batches to reduce this.

\textbf{What a storage-only client sees.} A client with only a sync secret
sees the shape of the tree and the roles of entries. It also sees growth
over time, key rotations, the anchors, whether inventories are equal,
and the parties that it exchanges cards with. Key rotations can show
when membership changes. It does not see authors, beneficiaries or
content.

\textbf{Keys can be copied.} A legitimate reader can copy plaintext. A
holder can give away its secrets. No protocol can prevent this. A copied
write or share seed does not let a party without a grant make a valid
entry. But if an attacker steals both a persona key and the secrets, the
attacker can act as that persona.

\textbf{No forward secrecy inside one key ring.} A party that holds the
read material for a position can read all content at and below that
position in its key ring, old and new. Only a key rotation
protects new content, and only from parties that do not get the new
keys.

\textbf{Nothing is global.} There is no global revocation, no remote
deletion, no invalidation of old entries and no winning branch.

\textbf{Spam.} A reusable invitation can receive unwanted claims. The
claims stay within the invitation's limits and the client's limits, and
they wait for the persona's attention.

\textbf{Secrets at rest.} An implementation must store each secret only
in encrypted form. The key for that encryption must not be in the same
store.

\textbf{Tests are not proof.} The specification says that passing the
test vectors is necessary for a conformance claim but is not enough. The
project's own measurements show this. \emph{Mutation testing} is a test
of the tests: a tester breaks the code on purpose and looks for a test
that fails. It found five blind spots in a suite that passed 102 of 102
records. Test records also cannot detect a race between two requests at
the same time. Review must cover these.

\section{Status}\label{sec:16}

The CoordX Protocol is a release candidate. It is not ready for production use. The
specification sets six gates before a candidate becomes a conformance
target:

\begin{enumerate}
\item
  The specification and its standards are fixed by digest.
\item
  Two independent generators make identical test vectors.
\item
  Two independent client implementations pass the client tests, and they
  work together through an independent relay that passes the relay
  tests.
\item
  An inference or machine-checked model checks the core model and authority rules.
\item
  Two independent attack reviews find no open critical problem.
\item
  The protocol author approves the candidate and its limits.
\end{enumerate}

The current text has 26 recorded changes since the earlier frozen
version. No question is open in the specification. One item is still
unresolved, and two are marked for the specification authors to look
at. This paper will change when the specification changes.

\section{Conclusion}\label{sec:17}

We have described a protocol that lets parties work on shared records
with no central server. A record is a tree of immutable, signed entries.
Each entry names the permission that allowed it, and any party can check
that permission from positive evidence alone. Permission needs a grant,
a valid identity and the matching secret together, so a copied key or a
lone grant is not enough. Content is encrypted, and servers carry only
sealed bytes. Clients copy frames by push and repair losses with keyed
inventory cards, so honest clients collect the same frames through any
servers. The protocol does not order entries, does not choose a winner
and does not need global agreement. It also states its limits directly:
nothing can be recalled, and a key rotation excludes a party only from
new content. Applications build their own views on top of these facts.

\appendix

\section{Terms}\label{app:A}

Each term has one meaning in this paper. The clause in brackets is where
the specification defines it.

\begin{description}
\item[AID] Autonomic identifier. The hash of an identity's root data. It names one identity for its whole life, with no outside authority (I2).
\item[anchors] The three public values of a key ring: a hash of the read root and the public keys of the write and share seeds (W2).
\item[authorised] Has a valid identity and a valid grant source for the kind, key ring and position (M5).
\item[capable] Authorised, and also holds the matching secret (M5).
\item[card] A sealed message with one keyed inventory row for each descriptor that two parties share (R3).
\item[certificate] The sealed part of a frame that names the author, the source rows and the claim (W4).
\item[challenge] A random value for one use, from a relay, that a inbox owner signs (R6).
\item[claim] A signed answer to an invitation (T3).
\item[connected] A client holds the other persona's public part and an address for it (T4).
\item[controller] An identity entry below the root. It can issue personas (I2).
\item[counterparty] Another persona that a persona is connected to (T4).
\item[delivery] A signed, sealed message that carries keys, sync secrets or evidence (T5).
\item[descriptor] The three values (object, key ring, scope position) that name one part of an object for synchronisation (M6).
\item[endpoint] The three values (AID, controller, persona) that name who acts (M2).
\item[entry] One item in an object. It never changes (C3).
\item[evidence item] A frame, certificate disclosure, identity envelope or root data, named by its hash (W6).
\item[frame] The exact bytes of one ordinary entry (W1).
\item[frontier] The members of a descriptor that have no member child at one client (R3).
\item[grant] An entry that gives one endpoint one or more kinds at a position (M4).
\item[grant source] A fact that gives permission: the signed origin claim of a root, a key rotation or a persona, or a grant (M4).
\item[invitation] A signed message that starts first contact (T2).
\item[judgment] A check that gives one exact result, such as \texttt{admitted} or \texttt{pending} (M7).
\item[key ring] One read root, one write seed and one share seed for one part of an object (M3).
\item[key rotation] An entry that starts a new key ring at a position (D4).
\item[kind] One of \texttt{read}, \texttt{write} or \texttt{share} (W1).
\item[inbox] A queue on a relay where others push packets for one persona (T1).
\item[material] The secret values that a client holds and uses, such as a read key or a seed (M5).
\item[client] Software that holds a user's keys and data and follows the protocol for that user (V1).
\item[object] A tree of entries that starts at one root (M1).
\item[packet] One sealed message that travels through a inbox (T1).
\item[persona] An identity entry that a controller issues. A persona authors entries (I2).
\item[position] A point in an object's tree: a root, a content entry or a persona. Content, grants and key rotations attach to a position (M1, I4).
\item[public part] The read root and sync root of a persona's own key ring, which the persona gives to others. It holds no write or share seed (I4, T2).
\item[relay] A server that holds inboxes. It sees no content (R6, S1).
\item[salt] A random value in each entry that makes its read key unique (W2).
\item[stage] One step in the fixed order of checks (M7).
\item[suite] A named set of cryptographic algorithms. The specification defines suite 0 (N2 to N6).
\item[sync secret] The secret that lets a holder synchronise one descriptor (R1).
\item[threshold] The number of keys in a set that must sign (I2).
\item[undertaking] A signed promise to store the members of a descriptor, within stated limits (T6).
\end{description}

\section{Specification listing}\label{app:B}

This appendix lists every clause of the specification. Each entry gives
the purpose of the clause in short words. Byte layouts are copied from
the specification. The specification holds the full rules, and it is the
only authority.

In layouts, \texttt{‖} joins bytes, \texttt{u8} to \texttt{u64} are
unsigned big-endian integers, and \texttt{lp(x)} is a 4-byte length
followed by \texttt{x}. \texttt{H(tag,\ m)} is BLAKE2b with a 32-byte
result and a purpose tag of at most 16 bytes. \texttt{KDF(tag,\ secret,\ info)} is
keyed BLAKE2b with a purpose tag.

The words MUST, SHOULD and MAY have the meanings of RFC 2119~\cite{rfc2119}.

\subsection{Protocol contract (C1 to
C8)}

These clauses bind the whole protocol. A later clause cannot make one
weaker.

\begin{itemize}
\item
  \textbf{C1. Complete normative authority.} The specification and the standards
  it names are the only source of meaning. Test vectors are evidence,
  not rules.
\item
  \textbf{C2. User authorship.} Exactly one user identity authors each
  entry. A client, relay or provider is never an author or an authority
  source.
\item
  \textbf{C3. Causal objects, not state.} An object holds only immutable
  entries and their single parent links. No rule depends on an entry
  being last, current or winning. Two entries with one parent are not a
  conflict.
\item
  \textbf{C4. Positive positional capability.} To act, an endpoint needs
  a valid grant source, a covering position, a grant for the key ring,
  proof of identity and the matching secret, all at once. Only positive
  evidence gives authority.
\item
  \textbf{C5. Determinism and privacy.} The same inputs always give one
  result. Arrival order and duplicates do not matter. A missing
  dependency can complete a \texttt{pending} check. The certificate is
  shown only to selected parties.
\item
  \textbf{C6. Topology-independent replication.} Clients that hold one
  descriptor and its sync secret collect the union of the retained
  frames, through any path. Proof closure travels only in sealed packets
  to a persona.
\item
  \textbf{C7. Forward-only exclusion.} Nothing recalls a key, seed,
  entry or plaintext. A key rotation excludes a holder only from new
  material.
\item
  \textbf{C8. Finite work and honest claims.} Everything that an
  attacker can start has a limit and an overload result. A limit never
  makes old or deep content invalid. A signature proves only its exact
  signed fact. A client keeps no frame that waits on a missing parent or
  key ring.
\end{itemize}

\subsection{Model (M1 to M8)}

\begin{itemize}
\item
  \textbf{M1. Content and capability entries.} A root and each content
  entry are content positions. Grants and key rotations attach to a
  content position as side branches. \texttt{covers(P,Q)} is true when P
  is Q or P is a content ancestor of Q.
\item
  \textbf{M2. Identity endpoints.} An endpoint is (AID, controller
  record identifier, persona record identifier). Several can be valid.
  None is current.
\end{itemize}

\begin{Verbatim}[fontsize=\footnotesize, xleftmargin=2em]
endpoint_ref =
    aid:bytes[32]
    ‖ controller_record_id:bytes[32]
    ‖ persona_record_id:bytes[32]
\end{Verbatim}

\begin{itemize}
\item
  \textbf{M3. Key rings.} A root, a key rotation and a persona each
  start a key ring: one read root, one write seed, one share seed. Read
  secrets go down content links. Write and share seeds cover the whole
  key ring.
\item
  \textbf{M4. Grant sources.} A grant source is the origin claim of a
  root, key rotation or persona, or a grant entry. A grant names one
  endpoint, a non-empty set of kinds and the exact placement key ring.
\item
  \textbf{M5. Capability.} \texttt{authorised} needs a valid identity, a
  valid source for that endpoint, the kind, the key ring and covering
  scope. \texttt{capable} also needs the matching material. Kinds have
  no order. Material is installed only when the client explicitly accepts
  it, or when the client makes the key ring itself.
\end{itemize}

\begin{Verbatim}[fontsize=\footnotesize, xleftmargin=2em]
authorised(A,k,Q,X,S) =
    valid_identity(A)
    and valid_grant_source(S)
    and beneficiary(S) = A
    and k in kinds(S)
    and key_ring(S) = X
    and key_ring_covers(X,Q)
    and covers(scope(S),Q)

capable(A,k,Q,X,S) =
    authorised(A,k,Q,X,S)
    and matching_material(A,k,Q,X,S)
\end{Verbatim}

\begin{itemize}
\item
  \textbf{M6. Descriptors and synchronisation.} A descriptor is (object,
  key ring, scope position). The key ring must cover the scope. A sync
  secret gives no identity, grant or storage promise.
\end{itemize}

\begin{Verbatim}[fontsize=\footnotesize, xleftmargin=2em]
descriptor_wire =
    object_id:bytes[32]
    ‖ key_ring_id:bytes[32]
    ‖ scope_position:bytes[32]
\end{Verbatim}

\begin{itemize}
\item
  \textbf{M7. Results.} Defines the result names of each judgment
  (public admission, identity proof, authority source, protected
  content, installed material) and the eight-stage order of checks. The
  first stage that decides gives the result. Missing items are found one
  layer at a time and listed up to 64.
\item
  \textbf{M8. Custody and durable authorship.} A secret set is judged
  separately as recoverable or not, and erasable or not. Before a client
  sends an artifact, it commits all its parts in one durable
  transaction.
\end{itemize}

\subsection{Suite 0 and encoding (N1 to
N6)}

\begin{itemize}
\item
  \textbf{N1. Notation.} Defines byte joining, integer sizes, length
  prefixes and parsing rules. A parser rejects overflow, wrong order,
  duplicates and trailing bytes.
\item
  \textbf{N2. Hashing and key derivation.} BLAKE2b~\cite{rfc7693} with a 32-byte result and a purpose tag of at most 16 bytes.
\end{itemize}

\begin{Verbatim}[fontsize=\footnotesize, xleftmargin=2em]
H(tag,m) =
    BLAKE2b(data=m, digest_size=32, personal=person(tag))

KDF(tag,secret,info) =
    BLAKE2b(data=info, key=secret,
            digest_size=32, personal=person(tag))
\end{Verbatim}

\begin{itemize}
\item
  \textbf{N3. Signatures.} Strict Ed25519 (RFC 8032)~\cite{rfc8032}. Points and
  \texttt{S} must be canonical. Small-order points are rejected. The
  check uses the cofactorless equation.
\item
  \textbf{N4. Key agreement, sealing and opening.} X25519 (RFC 7748)~\cite{rfc7748} with strict input checks, and XSalsa20-Poly1305~\cite{salsa,nacl,rfc8439}. A box is
  authenticated before any plaintext is released. Trials of keys use a
  fixed set of candidates. One success continues. More than one is
  \texttt{defective(non-unique-opening)}.
\end{itemize}

\begin{Verbatim}[fontsize=\footnotesize, xleftmargin=2em]
symmetric_seal(key,plaintext) =
    nonce:bytes[24] ‖ SB(key,nonce,plaintext)

public_seal(recipient_pk,plaintext) =
    ephemeral_pk:bytes[32]
    ‖ nonce:bytes[24]
    ‖ SB(KDF("kexseal",X25519(ephemeral_sk,recipient_pk),
              ephemeral_pk ‖ recipient_pk),
          nonce,plaintext)
\end{Verbatim}

\begin{itemize}
\item
  \textbf{N5. Freshness and bounds.} Random values come from a secure
  source. An author must not reuse a salt for two children of one
  parent. The suite-0 manifest and its identifier are fixed:
\end{itemize}

\begin{Verbatim}[fontsize=\footnotesize, xleftmargin=2em]
derive blake2b-256
sign ed25519-strict
seal xsalsa20poly1305
kex x25519
bounds 262144-1048576-1024-256-64-16

suite_id = H("suite",lp(manifest))
         = f11cdf4a758d40d9b3f76b3dc7ad3be3
           42186bc24fc28acc256f50511f2bb6d6
\end{Verbatim}

\begin{itemize}
\item
  \textbf{N6. Incorporated standards.} BLAKE2, Ed25519, X25519,
  XSalsa20/20, NaCl SecretBox, Poly1305, base64url~\cite{rfc4648}, JSON~\cite{rfc8259} and HTTPS origins~\cite{rfc6454}, with stricter encodings where the specification says so.
\end{itemize}

\subsection{Identity objects (I1 to
I6)}

\begin{itemize}
\item
  \textbf{I1. The identity object and its schema.} An identity is a tree
  of \texttt{root}, \texttt{controller} and \texttt{persona} entries in
  envelopes. Entries below a persona are ordinary frames. An unknown
  schema gives \texttt{opaque(schema)}.
\item
  \textbf{I2. Identity entries.} Defines the payloads, identifiers,
  envelope and signing strings. A threshold set has 1 to 16 unique,
  ordered keys. A root has at most 8 recovery sets. A client that holds a
  \texttt{valid} envelope must not replace it with another envelope for
  the same record.
\end{itemize}

\begin{Verbatim}[fontsize=\footnotesize, xleftmargin=2em]
authority_set =
    suite_id:bytes[32]
    ‖ threshold:u16
    ‖ n:u16
    ‖ n * signing_pk:bytes[32]

root_payload =
    lp("root")
    ‖ schema_id:bytes[32]
    ‖ root_authority:authority_set
    ‖ recovery_count:u16
    ‖ recovery_count * recovery_authority:authority_set

controller_payload =
    lp("controller")
    ‖ aid:bytes[32]
    ‖ root_ref:bytes[32]
    ‖ authoriser_kind:u8
    ‖ authoriser_index:u16
    ‖ controller_authority:authority_set
    ‖ permissions:u8

persona_payload =
    lp("persona")
    ‖ aid:bytes[32]
    ‖ controller_ref:bytes[32]
    ‖ persona_suite:bytes[32]
    ‖ persona_pk:bytes[32]
    ‖ powers:u8
    ‖ lp(author_target:bytes[0|96])
    ‖ anchors:bytes[96]

AID = H("aid", lp(root_payload))
record_id(payload) = H("idrecord", lp(payload))
endpoint_id = H("endpoint", endpoint_ref)

identity_envelope =
    lp(payload)
    ‖ authoriser_count:u16
    ‖ authoriser_count * (key_index:u16 ‖ signature:bytes[64])
    ‖ possession_count:u16
    ‖ possession_count * (key_index:u16 ‖ signature:bytes[64])
\end{Verbatim}

\begin{itemize}
\item
  \textbf{I3. Coexistence and recovery.} All valid controllers and
  personas stand. A recovery set prepared in the root can authorise a
  controller with the same AID, also under another suite.
\item
  \textbf{I4. The persona position.} A persona is a content position
  with its own key ring, its public part. Frames below it use the
  identity root's record identifier as their object identifier.
\item
  \textbf{I5. Addresses.} An address entry lists 1 to 8 inboxes and a
  delivery key. An annul entry cancels one address entry. A reader uses
  the valid, non-annulled entry with the greatest counter, and the
  smaller identifier on a tie.
\end{itemize}

\begin{Verbatim}[fontsize=\footnotesize, xleftmargin=2em]
address_payload =
    lp("address")
    ‖ counter:u64
    ‖ delivery_suite:bytes[32]
    ‖ delivery_pk:bytes[32]
    ‖ inbox_count:u8
    ‖ inbox_count * lp(inbox)

inbox =
    lp(binding)
    ‖ lp(locator)
    ‖ tag:bytes[32]
    ‖ kind:u8

annul_payload =
    lp("annul")
    ‖ address_entry_id:bytes[32]
\end{Verbatim}

\begin{itemize}
\item
  \textbf{I6. Disclosure and publication.} An identity is shown only to
  selected parties. Publication is opt-in: a persona with
  \texttt{connect} can sign an invitation, and a persona whose
  controller has \texttt{publish} can register a name for it.
\end{itemize}

\subsection{Ordinary frames (W1 to W6)}

\begin{itemize}
\item
  \textbf{W1. Frame format.} One format, \texttt{0x01}. Four roles:
  root, content, grant, key rotation. Kinds are \texttt{read=0x01},
  \texttt{write=0x02}, \texttt{share=0x04}. The entry identifier is the
  hash of the whole frame. An unknown format or role is
  \texttt{refused}. An unknown value inside a sealed box is
  \texttt{defective}.
\end{itemize}

\begin{Verbatim}[fontsize=\footnotesize, xleftmargin=2em]
anchors =
    read_commit:bytes[32]
    ‖ write_pk:bytes[32]
    ‖ share_pk:bytes[32]

ordinary_frame =
    format:u8
    ‖ suite_id:bytes[32]
    ‖ role:u8
    ‖ lp(object_id:bytes[0|32])
    ‖ lp(parent_id:bytes[0|32])
    ‖ lp(placement_key_ring:bytes[0|32])
    ‖ salt:bytes[32]
    ‖ position_key_commit:bytes[32]
    ‖ place_pk:bytes[32]
    ‖ lp(fresh_anchors:bytes[0|96])
    ‖ main_nonce:bytes[24]
    ‖ lp(main_box)
    ‖ certificate_nonce:bytes[24]
    ‖ place_signature:bytes[64]
    ‖ certificate_commit:bytes[32]
    ‖ material_use_signature:bytes[64]
    ‖ lp(certificate_box)

entry_id(F) = H("entry",F)
object_id(root F) = entry_id(F)
key_ring_id(root F)     = object_id(F)
key_ring_id(rotation F) = entry_id(F)
\end{Verbatim}

\begin{itemize}
\item
  \textbf{W2. Key ring anchors and read derivation.} Read secrets go
  down content links with the object, parent and salt. Grants and key
  rotations use a share-derived secret.
\end{itemize}

\begin{Verbatim}[fontsize=\footnotesize, xleftmargin=2em]
anchors(X) =
    H("readcommit",rX)
    ‖ public_key(Ed25519Seed(wX))
    ‖ public_key(Ed25519Seed(sX))

read_at(E,X) =
    KDF("readchild",read_at(parent(E),X),
        object_id(E) ‖ parent_id(E) ‖ salt(E))

entry_share(E,X) =
    KDF("shareentry",sX,
        object_id(E) ‖ parent_id(E) ‖ salt(E))
\end{Verbatim}

\begin{itemize}
\item
  \textbf{W3. Main box and placement signature.} The main box holds the
  payload type and payload. The material-use signature uses the write or
  share seed. The placement signature uses a one-time key and comes
  last. A client checks the placement signature first, at stage 1.
\end{itemize}

\begin{Verbatim}[fontsize=\footnotesize, xleftmargin=2em]
main_plaintext =
    payload_type:u8
    ‖ lp(payload)

main_key = KDF("content",frame_secret(E,X),
               lp(public_header))

material_use_preimage =
    "pk4.material-use:"
    ‖ lp(public_header)
    ‖ lp(main_box)
    ‖ certificate_nonce
    ‖ certificate_commit
    ‖ lp(certificate_box)

place_preimage =
    "pk4.place:"
    ‖ lp(public_header)
    ‖ lp(main_box)
    ‖ certificate_nonce
    ‖ certificate_commit
    ‖ material_use_signature
    ‖ lp(certificate_box)
\end{Verbatim}

Payload types are \texttt{application=0x00},
\texttt{retention-undertaking=0x06}, \texttt{retention-release=0x07},
\texttt{address=0x09} and \texttt{annul=0x0a}.

\begin{itemize}
\item
  \textbf{W4. Source rows and certificate.} The certificate names the
  author endpoint, one source row for each needed kind, and the role's
  claim. The persona signs it. It is sealed in the certificate box.
\end{itemize}

\begin{Verbatim}[fontsize=\footnotesize, xleftmargin=2em]
source_row =
    kind:u8
    ‖ source_entry_id:bytes[32]

certificate_core =
    aid:bytes[32]
    ‖ controller_record_id:bytes[32]
    ‖ persona_record_id:bytes[32]
    ‖ source_count:u8
    ‖ source_count * source_row
    ‖ lp(claim)

use_preimage =
    "pk4.use:"
    ‖ lp(public_header)
    ‖ lp(main_box)
    ‖ certificate_nonce
    ‖ lp(certificate_core)

certificate =
    lp(certificate_core)
    ‖ persona_signature:bytes[64]

certificate_commit = H("certificate",lp(certificate))
\end{Verbatim}

\begin{itemize}
\item
  \textbf{W5. What the certificate binds.} The persona signature binds
  the exact ciphertext, position, key ring, one-time key, sources and
  claim. A copy of the exact frame is the same entry. The public frame
  does not show the author.
\item
  \textbf{W6. Evidence items and proof closure.} Four evidence kinds,
  each named by its hash. A bundle has at most 64 items, in a fixed
  order. There is no limit on ancestry depth.
\end{itemize}

\begin{Verbatim}[fontsize=\footnotesize, xleftmargin=2em]
certificate_disclosure =
    entry_id:bytes[32]
    ‖ lp(certificate)

evidence_item =
    evidence_kind:u8
    ‖ if evidence_kind == 0x00: lp(ordinary_frame)
      if evidence_kind == 0x01: lp(certificate_disclosure)
      if evidence_kind == 0x02: lp(identity_envelope)
      if evidence_kind == 0x03: lp(record_core)
\end{Verbatim}

\subsection{Authority and validity (D1 to
D9)}

\begin{itemize}
\item
  \textbf{D1. Root origin.} A root's origin claim is valid when its
  framing, certificate, identity proofs, claim, anchors and signatures
  verify.
\item
  \textbf{D2. Grant.} Eight conditions: framing, a content parent, a
  covering key ring, one author and beneficiary, a valid share source, a
  valid source for each given kind, the share material-use signature,
  and valid dependencies.
\item
  \textbf{D3. Content entry.} Seven conditions: exact structure, a
  content parent, a covering key ring, a valid author identity, valid
  read and write sources, boxes that open, and valid signatures. The
  check reads no child, sibling or later entry.
\item
  \textbf{D4. Key rotation.} Seven conditions, with all three kinds from
  the old key ring. It starts a new key ring. Nobody is capable in it
  without the new material. Several key rotations are independent.
\item
  \textbf{D5. The authority-source judgment.} Decides if a grant source
  is valid from the exact certificate. A cycle, a wrong object or key
  ring, an uncovered position, a wrong beneficiary or a bad signature is
  \texttt{defective}. A missing item is \texttt{pending}.
\item
  \textbf{D6. Required capability.} Lists the kinds that each operation
  needs. To synchronise needs only a sync secret, with no identity
  capability.
\item
  \textbf{D7. No later repair.} A later grant cannot make an earlier
  entry valid. No rule looks for the absence of a revocation.
\item
  \textbf{D8. No revocation.} There is no revoke entry. Exclusion uses a
  key rotation. Old grants and material stay usable in the old key ring.
\item
  \textbf{D9. Public admission.} Checks bytes, identifiers, the parent
  and key ring, membership, both public signatures and limits. A client
  must hold the placement key ring, admitted, before it accepts a frame.
  A frame that waits is not kept.
\end{itemize}

\subsection{First contact, delivery and retention (T1 to
T6)}

\begin{itemize}
\item
  \textbf{T1. Inbox addresses.} An address is transport advice only. A
  push goes to the first inbox that accepts it, in order. The sender
  retries with a doubling wait and seals again each time. A packet whose
  every inbox is unusable is removed, except a claim and a T4 card. A
  packet stays queued when the client cannot read the receiver's address.
  One-shot packets and packs take turns, so neither waits long behind
  the other.
\item
  \textbf{T2. Invitation.} A signed message that carries the inviter's
  public part and a claim key. \texttt{max\_claims} is 1 for private and
  2 to 65535 for reusable.
\end{itemize}

\begin{Verbatim}[fontsize=\footnotesize, xleftmargin=2em]
public_part =
    object_id:bytes[32]
    ‖ key_ring_id:bytes[32]
    ‖ read_root:bytes[32]
    ‖ sync_root:bytes[32]

invitation_core =
    version:u8
    ‖ invitation_secret:bytes[32]
    ‖ inviter_endpoint:bytes[32]
    ‖ claim_suite:bytes[32]
    ‖ claim_pk:bytes[32]
    ‖ max_claim_bytes:u32
    ‖ max_claims:u16
    ‖ inbox_count:u8
    ‖ inbox_count * lp(inbox)
    ‖ public_part
    ‖ lp(display:bytes[0..256])
    ‖ evidence_count:u8
    ‖ evidence_count * lp(evidence_item)

invitation =
    lp(invitation_core)
    ‖ inviter_signature:bytes[64]

invitation_id  = H("invitation", invitation)
rendezvous_id  = H("rendezvous", invitation_id)
rendezvous_key = KDF("rendezvous", invitation_secret, rendezvous_id)
\end{Verbatim}

\begin{itemize}
\item
  \textbf{T3. Claim.} A signed answer with the claimant's public part
  and a fresh reply key, sealed twice and pushed one inbox at a time.
\end{itemize}

\begin{Verbatim}[fontsize=\footnotesize, xleftmargin=2em]
claim_core =
    version:u8
    ‖ invitation_id:bytes[32]
    ‖ target_endpoint:bytes[32]
    ‖ claimant_endpoint:bytes[32]
    ‖ reply_suite:bytes[32]
    ‖ reply_pk:bytes[32]
    ‖ nonce:bytes[32]
    ‖ inbox_count:u8
    ‖ inbox_count * lp(inbox)
    ‖ public_part
    ‖ lp(note:bytes[0..1024])
    ‖ evidence_count:u8
    ‖ evidence_count * lp(evidence_item)

claim =
    lp(claim_core)
    ‖ claimant_signature:bytes[64]

claim_id = H("claim", claim)

claim_packet =
    rendezvous_id
    ‖ claim_id
    ‖ symmetric_seal(rendezvous_key,
        public_seal(claim_pk, claim))
\end{Verbatim}

\begin{itemize}
\item
  \textbf{T4. Connection.} The inviter accepts the claim, installs the
  claimant's public part and sends one card to the reply key. Connection
  is only mutual public parts and addresses. A client can detach a party
  that floods it. A claimant must give up after a local limit.
\item
  \textbf{T5. Delivery.} A signed message, sealed to the recipient's
  delivery key, with 1 to 64 items. No client installs an item without
  explicit acceptance. Also defines the material request and the
  evidence request.
\end{itemize}

\begin{Verbatim}[fontsize=\footnotesize, xleftmargin=2em]
delivery_core =
    version:u8
    ‖ sender_endpoint:bytes[32]
    ‖ recipient_endpoint:bytes[32]
    ‖ item_count:u8
    ‖ item_count * lp(delivery_item)

delivery =
    lp(delivery_core)
    ‖ sender_signature:bytes[64]

delivery_packet =
    public_seal(delivery_pk, delivery)

delivery_item =
    if item_kind == 0x00:
        item_kind:u8
        ‖ target:descriptor
        ‖ grant_ref:bytes[32]
        ‖ kinds:u8
        ‖ read_seed:bytes[32]
        ‖ write_seed:bytes[32]
        ‖ share_seed:bytes[32]
        ‖ evidence_count:u8
        ‖ evidence_count * lp(evidence_item)
    if item_kind == 0x01:
        item_kind:u8
        ‖ target:descriptor
        ‖ scope_sync_secret:bytes[32]
    if item_kind == 0x02:
        item_kind:u8
        ‖ evidence_count:u8
        ‖ evidence_count * lp(evidence_item)

material_request =
    version:u8
    ‖ requester_endpoint:bytes[32]
    ‖ grant_ref:bytes[32]
    ‖ target:descriptor

evidence_request =
    version:u8
    ‖ requester_endpoint:bytes[32]
    ‖ target_entry_id:bytes[32]
    ‖ dependency_count:u8
    ‖ dependency_count * dependency_ref

dependency_ref =
    dependency_kind:u8
    ‖ dependency_id:bytes[32]
\end{Verbatim}

The item kinds are a capability item (\texttt{0x00}), a sync item
(\texttt{0x01}) and an evidence item (\texttt{0x02}). Dependency kinds
are \texttt{frame=0x00}, \texttt{certificate=0x01},
\texttt{identity-record=0x02} and \texttt{endpoint=0x03}.

\begin{itemize}
\item
  \textbf{T6. Retention undertakings.} The only protocol promise to
  store. It is a valid content entry in a branch of the author's
  identity, read by one appointer. It holds limits and a snapshot. It
  lasts until a release.
\end{itemize}

\begin{Verbatim}[fontsize=\footnotesize, xleftmargin=2em]
retention_undertaking =
    target:descriptor
    ‖ max_frames:u64
    ‖ max_bytes:u64
    ‖ snapshot_count:u64
    ‖ snapshot_bytes:u64
    ‖ snapshot_digest:bytes[32]

retention_release =
    undertaking_entry_id:bytes[32]
\end{Verbatim}

\subsection{Synchronisation and inboxes (R1 to
R6)}

\begin{itemize}
\item
  \textbf{R1. Sync roots and keys.} Each root and key rotation makes a
  sync root. Sync values go down content links. The sync key seals sync
  messages and keys card rows. A client never gives a sync secret to a
  relay.
\end{itemize}

\begin{Verbatim}[fontsize=\footnotesize, xleftmargin=2em]
sync_at(E,X) =
    KDF("syncchild",sync_at(parent(E),X),
        object_id(E) ‖ parent_id(E) ‖ salt(E))

sync_key(D,syncP) =
    KDF("synckey",syncP,encode(D))
\end{Verbatim}

\begin{itemize}
\item
  \textbf{R2. Descriptor membership.} Members are the root (when the
  descriptor is the root's), and the content, grants and key rotations
  of that key ring that the scope covers. Parent frames outside that set
  are dependencies, not members.
\item
  \textbf{R3. Frontier and messages.} Defines the frontier, the
  inventory digest, the card row, the four sync messages and the card. A
  bad sync message is \texttt{discarded} with no answer. A card row
  verifies only when it equals the holder's own row.
\end{itemize}

\begin{Verbatim}[fontsize=\footnotesize, xleftmargin=2em]
frontier(D) =
    { E in members(D) :
      no F in members(D) has parent(F) = E }

inventory_digest(D) =
    H("inventory",
      descriptor_wire(D) ‖ u64(n) ‖ i1 ‖ ... ‖ in)

card_row(D) =
    descriptor_wire(D)
    ‖ u64(n)
    ‖ KDF("cardrow", sync_key(D), inventory_digest(D))

sync_plaintext =
    version:u8
    ‖ descriptor_wire
    ‖ correlation:bytes[16]
    ‖ kind:u8
    ‖ lp(body)

pack_item =
    requested:u8
    ‖ entry_id:bytes[32]
    ‖ lp(ordinary_frame)

card =
    version:u8
    ‖ correlation:bytes[16]
    ‖ count:u16
    ‖ count * card_row
\end{Verbatim}

\begin{table}[htbp]
\caption{Sync message kinds (R3)}\label{tab:3}
\centering\scriptsize
\begin{tabular}{@{}>{\raggedright\arraybackslash}p{0.21\linewidth}>{\raggedright\arraybackslash}p{0.74\linewidth}@{}}
\hline
\textbf{kind} & \textbf{body} \\ \hline
\texttt{leaf-request=0x01} & \texttt{requester\_endpoint:bytes[32] ‖ lp(after\_id:bytes[0|32]) ‖ limit:u16} \\
\texttt{leaf-page=0x02} & \texttt{card\_row ‖ lp(after\_id:bytes[0|32]) ‖ returned:u16 ‖ returned * entry\_id:bytes[32] ‖ more:u8} \\
\texttt{want=0x03} & \texttt{requester\_endpoint:bytes[32] ‖ count:u16 ‖ count * entry\_id:bytes[32]} \\
\texttt{pack=0x04} & \texttt{card\_row ‖ count:u16 ‖ count * pack\_item} \\
\hline
\end{tabular}
\end{table}

The \texttt{sync\_plaintext} version is one. A card's version is zero.

\begin{itemize}
\item
  \textbf{R4. Reconciliation.} Frames move by push to known holders.
  Cards find differences. Exchanges repair them. There is no
  acknowledgement. Any holder can repair any other holder. A client knows
  the sender of a sync message only in three defined cases.
\item
  \textbf{R5. Removed.} The number stays so that later numbers do not
  change.
\item
  \textbf{R6. The \texttt{https-mailbox} binding.} Strict JSON over
  HTTPS. Eight operations. \texttt{packet\_bytes} is 1048576 on every
  relay. A challenge expires at its deadline. Only release removes
  packets. The relay keeps no receipt, sender identity, object or
  content.
\end{itemize}

\begin{table}[htbp]
\caption{Operations of the \texttt{https-mailbox} binding (R6)}\label{tab:4}
\centering\footnotesize
\begin{tabular}{@{}>{\raggedright\arraybackslash}p{0.20\linewidth}>{\raggedright\arraybackslash}p{0.34\linewidth}>{\raggedright\arraybackslash}p{0.36\linewidth}@{}}
\hline
\textbf{operation} & \textbf{request} & \textbf{success} \\ \hline
declare & \texttt{GET /mailbox} & \texttt{200} with the relay's limits and kinds \\
register & \texttt{PUT /mailbox/<tag>} & \texttt{201}, or \texttt{200} when already registered identically \\
withdraw & \texttt{DELETE /mailbox/<tag>} & \texttt{204} \\
push & \texttt{POST /mailbox/<tag>} & \texttt{202} \\
open & \texttt{POST /session} & \texttt{201} with a handle \\
pull & \texttt{GET /pull} & one \texttt{200 text/event-stream} \\
release & \texttt{POST /release} & \texttt{204} \\
serve invitation & \texttt{GET /invite/<name>} & \texttt{200} with the invitation \\
\hline
\end{tabular}
\end{table}

\subsection{Required client behaviour (Q1 to
Q6)}

\begin{itemize}
\item
  \textbf{Q1. Operations and results.} A client must do nine groups of
  operations, from identity creation to erasure. Each operation answers
  \texttt{committed}, \texttt{pending}, \texttt{refused-by-evidence} or
  \texttt{local-failure}, with fixed rules to combine several answers.
\item
  \textbf{Q2. Observations.} The exact facts that a client shows about
  identifiers, for example admission results, capability and queue
  counts.
\item
  \textbf{Q3. Change hints.} A signal that something changed. A hint is
  never evidence.
\item
  \textbf{Q4. Export, custody transfer and migration.} Export gives
  readable payloads with their judgments. Custody transfer copies a
  selected complete closure to another conformant client. An import with any frame that is
  not admitted installs nothing.
\item
  \textbf{Q5. Crash-monotone authoring.} Random values are taken once.
  The artifact is committed before it is first sent. A retry sends the
  same bytes.
\item
  \textbf{Q6. Erasure.} Six steps in order: record, prepare, release
  undertakings, destroy every copy, record completion, report.
\end{itemize}

\subsection{Security contract and limits (S1 to
S8)}

\begin{itemize}
\item
  \textbf{S1. What a relay and a sync-only client see.} Lists the exact
  facts that each one sees and does not see.
\item
  \textbf{S2. Sync secrets.} A sync secret allows listing and sending,
  not reading, writing or sharing. A bad holder can use up bounded
  capacity.
\item
  \textbf{S3. Material.} Read material shows the content it covers.
  Copied write or share material cannot make a valid entry without a
  grant.
\item
  \textbf{S4. Immutability and key rotation.} No global revocation,
  remote deletion or winning branch. A key rotation protects only new
  content.
\item
  \textbf{S5. Identity.} Identities need no directory. Recovery keeps
  the AID. A stolen root cannot be repaired inside the protocol.
\item
  \textbf{S6. Retention and inboxes.} An undertaking is evidence of a
  promise, not of storage. A inbox makes no promise.
\item
  \textbf{S7. Suites.} A client can stop new use of a suite or condemn
  it. After it condemns a suite, it reports \texttt{opaque(suite)}. It does not change shared validity.
\item
  \textbf{S8. Implementation.} Constant-time secret operations, erasure
  of short-lived secrets, authentication before plaintext, and parsing
  separate from allocation. Stored secrets must be encrypted under a key
  held outside the store. Passing vectors alone is not enough.
\end{itemize}

\subsection{Conformance (V1 to V4)}

\begin{itemize}
\item
  \textbf{V1. Roles.} A claim is for one role: a client or a relay. A
  product can do both but tests them separately. No alternate form of an
  encoding, signature string or result is allowed.
\item
  \textbf{V2. Evidence.} Lists 55 client items and 11 relay items that the
  test evidence must cover.
\item
  \textbf{V3. Replication.} At least three independent clients, with
  different start data and changing relays, must reach equal frontiers.
\item
  \textbf{V4. Candidate to target.} The six gates in Section~\ref{sec:16}.
\end{itemize}

\subsection{Composition patterns
(informative)}

The specification gives 15 patterns as advice: application views,
detached content, collective parties, capability bundles, replicas and
progress, evidence availability, retention and validating custody, the
key rotation workflow, transport reachability, recovery custody,
personas and devices, connection readership, suite maintenance,
portability and SDKs, and external services.

\section{How to update this paper}\label{app:C}

When \texttt{SPEC.md} changes:

\begin{enumerate}
\item
  Put the new SHA-256 digest and profile name in the Source paragraph of this paper
  and in reference~\cite{spec}.
\item
  Compare the changed clauses with Appendix~\ref{app:B}, and update the summaries
  and byte layouts.
\item
  Check Sections~\ref{sec:3} to~\ref{sec:15} for any statement that the change affects.
\item
  Keep the style: short sentences, active voice, one idea in each
  sentence, and a definition for each new term in Appendix~\ref{app:A}.
\item
  Build the PDF again with \texttt{tectonic whitepaper.tex}.
\end{enumerate}

\end{document}